\documentclass[a4paper,dvips,12pt]{article}
\usepackage{amsmath}
\usepackage{epsfig,graphics,graphicx}

\input{tcilatex}

\begin{document}

\title{\textbf{Iterated Resolvent Function for the Ladder Bethe-Salpeter
Equation}}
\author{J.H.O. Sales$^{a,b}$, T.Frederico$^{c}$, B.M.Pimentel$^{d}$ and
B.V.Carlson$^{c}$. \\
%EndAName
$^{a}$Instituto de Ci\^{e}ncias Exatas, Universidade Federal de Itajub\'{a}, 
\\
37500-000, Itajub\'{a}-MG, Brazil. \\
$^{b}$Funda\c{c}\~{a}o de Ensino e Pesquisa de Itajub\'{a}, \\
37501-002, Itajub\'{a}-MG-Brazil.\\
$^{c}$Dep.de F\'{\i}sica, Instituto Tecnol\'{o}gico de Aeron\'{a}utica, \\
Centro T\'ecnico Aeroespacial, \\
12.228-900 S\~{a}o Jos\'{e} dos Campos, S\~{a}o Paulo, Brazil.\\
$^{d}$Instituto de F\'{\i}sica Te\'{o}rica-UNESP, 01405-900\\
S\~{a}o Paulo, Brazil.}
\date{\today}
\maketitle

\begin{abstract}
We develop the formal connection of the field theoretical Bethe-Salpeter
equation including the ladder approximation with its representation on the
light-front for a bosonic model. We use the light-front Green's function for
the N-particle system for two-particles plus N-2 intermediate bosons. We
derive an infinite set of coupled hierarchy equations or iterated resolvents
from which the Green's function can be calculated. These equations allow a
consistent truncation of the light-front Fock-space. We show explicitly the
represention of the covariant two-body T-matrix and bound state vertex in
the light-front.
\end{abstract}

%\twocolumn
%\narrowtext

\section{Introduction}

The Bethe-Salpeter (BS) equation provides a field-theoretical framework to
study two-particle bound states \cite{bs}. In general, for practical
applications its two-particle irreducible kernel is truncated in lowest
order. Nowadays it is possible to solve it including an infinite set of
ladder and crossed ladder diagrams in a nonperturbative field theoretical
calculation \cite{tjon}. Recently the ladder BS in Minkowiski space has also
been solved \cite{will97}.

In this sense the discussion of three-dimensional reductions of the BS
equation seems out of date \cite{brown}. However, much effort has been made
in using the three-dimensional reduction using light-front coordinates.
There is the hope that, by using light-front quantization, one could be able
to understand several aspects of low energy QCD without loosing the
kinematical boost invariance of the wave-function \cite{pe90}. In the past,
the concept of light-front wave-functions has also been applied in the
context of nuclear physics to describe the deuteron and its properties \cite%
{frank,karmanov}.

The representation of the ladder BS equation in the infinitum momentum frame
limit was derived by Weinberg \cite{wein}. However, the result corresponds
to the truncation of the intermediate light-front Fock-space to
three-particles. In this approximation numerical results were obtained in
different contexts, for bosonic models \cite{ji86,ji94} and fermionic models 
\cite{pe90}. An explicit systematic expansion of the covariant BS equation
in the light-front is still missing in the literature. It is our aim to fill
this formal gap. At the same time it is also necessary to perform detailed
numerical calculations comparing the covariant BS ladder approximation to
its truncation in the light-front Fock-space.

It is well known that light-front perturbation theory is equivalent to the
covariant perturbative expansion in field theory\cite{yan,bak}. Thus, in
principle, it is possible to find the exact representation of the BS
equation in the light-front. For this purpose we use a bosonic model for
which the interaction Lagrangean is defined by 
\begin{equation}
\mathcal{L}_{I}=g_{S}\phi _{1}^{\dagger }\phi _{1}\sigma +g_{S}\phi
_{2}^{\dagger }\phi _{2}\sigma ,
\end{equation}
where the bosons $\phi _{1}$ and $\phi _{2}$ have equal masses, $m$, the
intermediate boson, $\sigma $, has mass $\mu $ and $g_{S}$ is the coupling
constant.

Beginning from Dirac's idea \cite{di49} of representing the dynamics of the
quantum system at ligth-front times $x^{+}=t+z$, we derive the two-body
Green's function from the covariant propagator that evolves the system from
one light-front hyper-surface to another one. The light-front Green's
function is the probability amplitude for an initial state at $x^{+}=0$ do
evolvey to a final state in the Fock-state at some $x^{+}$, where the
evolution operator is defined by the light-front Hamiltonian \cite{nam}. The
two-body Green's function includes the propagation of intermediate states
with any number of particles. When it is conveniently constrained, it
results in the summation of the covariant ladder. The exact representation
of the covariant off-shell two-body amplitude in terms of the light-front
T-matrix is given here, as well as the light-front representation of the
four-dimensional vertex of the bound state.

The two-body light-front Green's function satisfies a hierarchy of coupled
equations which includes virtual intermediate propagation for any number of
particles. A consistent truncation can be performed, and in lowest order it
is related to the Weinberg equation.

This work is divided as follows sections. In section $II$, we present the
notation used through the work, discussing the one boson propagator in the
light-front. In section $III$, we introduce the two-body light-front Green's
function in perturbation theory, and explicitly evaluated it up to $%
g_{S}^{4} $, including one boson exchange and the box-diagram. Then, in
section $IV$, we generalize this discussion to any order in the ladder, and
present the coupled hierarchy equations which, if solved, give the
light-front two-body Green's function. The hierarchy of equations does not
include closed loops in the bosons $\Phi _{1}$ and $\Phi _{2}$. We relate
the off-shell two-body scattering amplitude to the light-front T-matrix. In
section $V$, we deduce the eigenvalue equation for the squared interacting
mass operator of the bound state. We express the four dimensional vertex in
terms of the light-front bound state wave-function. The approximation of the
ladder BS equation in the light-front with up to four particles in lowest
order in the kernel is given. Our conclusions are summarized in section $IV$.

\section{Notation}

The kinematics on the light-front are defined by the momentum canonically
conjugate to the light-front coordinates, $x^{-}=t-z$ and $\vec{x}_{\perp }$%
. The momentum $k^{+}=k^{0}+k^{3}$ is canonically conjugate to $x^{-}$ and $%
\vec{k}_{\perp }$ to $\vec{x}_{\perp }$. The eigenfunctions of the momentum
operators $k^{+}$ and $\vec{k_{\perp }}$ are defined by 
\begin{equation}
<x|k>\equiv <x^{-},\vec{x}_{\perp }|k^{+},\vec{k}_{\perp }>=e^{-i(\frac{1}{2}%
k^{+}x^{-}-\vec{k}_{\perp }.\vec{x}_{\perp })}.  \label{basis}
\end{equation}
The basis states are eigenfunctions of the free $k_{0}^{-}$ operator 
\begin{equation}
k_{0}^{-}|k>=\frac{k_{\perp }^{2}+m^{2}}{k^{+}}|k>.  \label{km}
\end{equation}
The states $|k>$ form an orthonormal and complete basis: 
\begin{equation}
\int \frac{dk^{+}d^{2}k_{\perp }}{2(2\pi )^{3}}<x^{\prime }|k><k|x>=\delta
(x^{\prime }{}^{+}-x^{+})\delta (\vec{x}_{\perp }^{\prime }-\vec{x}_{\perp
}).
\end{equation}

The free one-body Green's function for particle propagation is defined by
the operator 
\begin{equation}
G_{0}^{(1p)}(k^{-})=\frac{\theta (k^{+})}{k^{-}-k_{0}^{-}+i\varepsilon };
\label{g1p}
\end{equation}
and, for antiparticle propagation, is given by 
\begin{equation}
G_{0}^{(1a)}(k^{-})=\frac{\theta (-k^{+})}{k^{-}-k_{0}^{-}-i\varepsilon }.
\label{g1a}
\end{equation}

The function defined by Eq.(\ref{g1p}) is the Green's function of the
operator equation 
\begin{equation}
\left( k^{-}-k_{0}^{-}\right) \left(
G_{0}^{(1p)}(k^{-})+G_{0}^{(1a)}(k^{-})\right) =1.
\end{equation}

The Feynman propagator is defined by 
\begin{equation}
S^{(1)}(k^{\mu })=\frac{i}{k^{+}}G_{0}^{(1p)}-\frac{i}{|k^{+}|}G_{0}^{(1a)}=%
\frac{i}{k^{+}(k^{-}-\frac{k_{\perp }^{2}+m^{2}-i\varepsilon }{k^{+}})},
\label{s1c}
\end{equation}
where the phase-space factor $1/|k^{+}|$ has been introduced.

The Green's function of Eq.(\ref{g1p}) is the Fourier transform of the
single-boson propagator for forward direction in the light-front time while,
for antiparticle states, the propagation is backward in the light-front
time. The space-time propagator is given by: 
\begin{eqnarray}
\widetilde{S}^{(1)}(x^{+}) &=&\int \frac{dk^{-}}{2\pi }\frac{ie^{-\frac{i}{2}%
k^{-}x^{+}}}{k^{+}(k^{-}-\frac{k_{\perp }^{2}+m^{2}-i\varepsilon }{k^{+}})} 
\notag \\
&=&\frac{1}{k^{+}}e^{\frac{i}{2}\frac{k_{\perp }^{2}+m^{2}}{k^{+}}%
x^{+}}\theta (k^{+})\theta (x^{+})+\frac{1}{|k^{+}|}e^{\frac{i}{2}\frac{%
k_{\perp }^{2}+m^{2}}{k^{+}}x^{+}}\theta (-k^{+})\theta (-x^{+})\ .
\label{s1t}
\end{eqnarray}

\section{Two-body Green's Function}

\subsection{Free Green's Function}

The two-body Green's functions can be derived from the covariant propagator
for two particles propagating at equal light-front times. Without losing
generality, we are going to restrict our calculation to total momentum $%
K^{+} $ positive and the corresponding forward light-front time propagation.
In this case the propagator from $x^{+}=0$ to $x^{+}>0$ is given by: 
\begin{equation}
\widetilde{S}^{(2)}(x_{1}^{\prime \mu },x_{2}^{\prime \mu };x_{1}^{\mu
},x_{2}^{\mu })=\int \frac{d^{4}k_{1}}{\left( 2\pi \right) ^{4}}\frac{%
d^{4}k_{2}}{\left( 2\pi \right) ^{4}}\frac{ie^{-ik_{1}^{\mu }\left( x_{1\mu
}^{\prime }-x_{1\mu }\right) }}{k_{1}^{2}-m^{2}+i\varepsilon }\frac{%
ie^{-ik_{2}^{\mu }\left( x_{2\mu }^{\prime }-x_{2\mu }\right) }}{%
k_{2}^{2}-m^{2}+i\varepsilon }.
\end{equation}
At equal light-front times $x_{1}^{+}=x_{2}^{+}=0$ and $x^{\prime
+}{}_{1}=x^{\prime +}{}_{2}=x^{+}$, the propagator is written as: 
\begin{equation}
\widetilde{S}^{(2)}(x^{+})=\widetilde{S}_{1}^{(1)}(x^{+})\widetilde{S}%
_{2}^{(1)}(x^{+}),  \label{s2a}
\end{equation}
where the one-body propagators, $\widetilde{S}_{i}^{(1)}$, corresponding to
the light-front propagators of particles $i=1$ or $2$, are defined by Eq.(%
\ref{s1t}). We have explicitly: 
\begin{equation}
\widetilde{S}^{(2)}(x^{+})=-\int \frac{dk_{1}^{-}}{\left( 2\pi \right) }%
\frac{dk_{2}^{-}}{\left( 2\pi \right) }\frac{e^{-\frac{i}{2}k_{1}^{-}x^{+}}}{%
k_{1}^{+}\left( k_{1}^{-}-\frac{k_{1\perp }^{2}+m^{2}-i\varepsilon }{%
k_{1}^{+}}\right) }\frac{e^{-\frac{i}{2}k_{2}^{-}x^{+}}}{k_{2}^{+}\left(
k_{2}^{-}-\frac{k_{2\perp }^{2}+m^{2}-i\varepsilon }{k_{2}^{+}}\right) }.
\label{s2b}
\end{equation}

The Fourier transform to the total light-front energy $(K^{-})$ is given by 
\begin{equation}
{S}^{(2)}(K^{-}):=\frac{1}{2}\int dx^{+}e^{\frac{i}{2}K^{-}x^{+}}\widetilde{S%
}^{(2)}(x^{+})\ ,
\end{equation}
which result in 
\begin{equation}
{S}^{(2)}(K^{-})=-\frac{1}{\left( 2\pi \right) }\int \frac{dk_{1}^{-}}{%
k_{1}^{+}k_{2}^{+}}\frac{1}{\left( k_{1}^{-}-\frac{k_{1\perp
}^{2}+m^{2}-i\varepsilon }{k_{1}^{+}}\right) }\frac{1}{\left(
K^{-}-k_{1}^{-}-\frac{k_{2\perp }^{2}+m^{2}-i\varepsilon }{K^{+}-k_{1}^{+}}%
\right) }\ ,
\end{equation}
where $K^{-}=k_{1}^{-}+k_{2}^{-}$.

We perform the analytical integration in the $k_{1}^{-}$ momentum by
evaluating the residue at the pole $k_{1}^{-}=(k_{\perp
}^{2}+m^{2}-i\varepsilon )/k^{+}$. It implies that only $k_{1}^{+}$ in the
interval $0<k_{1}^{+}<K^{+}$ gives a nonvanishing contribution to the
integration. The result is 
\begin{equation}
{S}^{(2)}(K^{-})=\frac{\theta (k_{1}^{+})\theta (K^{+}-k_{1}^{+})}{%
k_{1}^{+}\left( K^{+}-k_{1}^{+}\right) }\frac{i}{\left(
K^{-}-K_{0}^{(2)-}+i\varepsilon \right) },  \label{s2km}
\end{equation}
where 
\begin{equation}
K_{0}^{(2)-}=\frac{k_{1\perp }^{2}+m^{2}}{k_{1}^{+}}+\frac{k_{2\perp
}^{2}+m^{2}}{K^{+}-k_{1}^{+}},  \label{k02}
\end{equation}
with $K_{0}^{(2)-}$ being the light-front Hamiltonian of the free
two-particle system. For $x^{+}<0$, $S^{(2)}(x^{+})\ =\ 0$ due to our choice
of $K^{+}>0$. Observe that $S^{(2)}(K^{-})$ is written in Eq.(\ref{s2km}) in
operator form with respect to $k^{+}$ and $\vec{k}_{\perp }$.

The free two-body Green's function is given by 
\begin{equation}
G_{0}^{(2)}(K^{-})=\theta (k_{1}^{+})\theta (K^{+}-k_{1}^{+})\frac{1}{\left(
K^{-}-K_{0}^{(2)-}+i\varepsilon \right) }\ .  \label{g02}
\end{equation}
The difference between the free two-body Green's function and $%
S^{(2)}(K^{-}) $, Eq.(\ref{s2km}), is the phase-space factor for particles 1
and 2.

The generalization for the $N$ particle system is 
\begin{equation}
{S}^{(N)}(K^{-})=\left[ \prod_{j=1}^{N}\frac{\theta (k_{j}^{+})\theta
(K^{+}-k_{j}^{+})}{k_{j}^{+}}\right] \frac{i}{\left(
K^{-}-K_{0}^{(N)-}+i\varepsilon \right) },  \label{g0n}
\end{equation}
where $K_{0}^{(N)-}$ is the free light-front Hamiltonian of the $N$-particle
system, which is given by 
\begin{equation}
K_{0}^{(N)-}=\sum_{j=1}^{N}\frac{k_{j\perp }^{2}+m_{j}^{2}}{k_{j}^{+}},
\label{k0n}
\end{equation}
with $k_{j}^{+}>0$ and $K^{+}=\sum_{j=1}^{N}k_{j}^{+}$. The many-body free
Green's function is given by 
\begin{equation}
G_{0}^{(N)}(K^{-})=\left[ \prod_{j=1}^{N}\theta (k_{j}^{+})\theta
(K^{+}-k_{j}^{+})\right] \frac{1}{\left( K^{-}-K_{0}^{(N)-}+i\varepsilon
\right) },
\end{equation}

In the following subsection, we will include the exchange of a boson between
the two-particles, and calculate the correction to the light-front
propagator.

\subsection{Green's Function $\mathcal{O}(g_S^2)$}

The perturbative correction to the two-body propagator which comes from the
exchange of one intermediate virtual boson, is given by 
\begin{equation}
\Delta \widetilde{S}_{g_{S}^{2}}^{(2)}(x^{+})=\left( \frac{ig_{S}}{2}\right)
^{2}\int d\overline{x}_{1}^{+}d\overline{x}_{2}^{+}\widetilde{S}_{1^{\prime
}}^{(1)}(x^{+}-\overline{x}_{1}^{+})\widetilde{S}_{2^{\prime }}^{(1)}(x^{+}-%
\overline{x}_{2}^{+})\widetilde{S}_{\sigma }^{(1)}(\overline{x}_{1}^{+}-%
\overline{x}_{2}^{+})\widetilde{S}_{1}^{(1)}(\overline{x}_{1}^{+})\widetilde{%
S}_{2}^{(1)}(\overline{x}_{2}^{+})\ .  \label{g2i}
\end{equation}
The intermediate boson, $\sigma $, propagates between the time interval $%
\overline{x}_{1}^{+}-\overline{x}_{2}^{+}$. The labels $1$ and $2$ and $%
1^{\prime }$ and $2^{\prime }$ in the particle propagators indicate initial
and final states, respectively.

Performing the Fourier transform from $x^{+}$ to $K^{-}$, for the total
kinematical momentum $K^{+}$, which we choose positive, and for $K_{\perp }$%
, we find 
\begin{eqnarray}
\Delta {S}_{g_{S}^{2}}^{(2)}(K^{-})=i\frac{(ig_{S})^{2}}{\left( 2\pi \right)
^{2}}\int &&\frac{dk_{1^{\prime }}^{-}dk_{1}^{-}}{k_{1^{\prime
}}^{+}(K^{+}-k_{1^{\prime }}^{+})k_{1}^{+}(K^{+}-k_{1}^{+})(k_{1^{\prime
}}^{+}-k_{1}^{+})}\frac{1}{\left( k_{1^{\prime }}^{-}-k_{1}^{-}-\frac{\left(
k_{1^{\prime }}-k_{1}\right) _{\perp }^{2}+m_{\sigma }^{2}-i\varepsilon }{%
k_{1^{\prime }}^{+}-k_{1}^{+}}\right) }\times  \notag \\
&&\frac{1}{\left( k_{1^{\prime }}^{-}-\frac{k_{1^{\prime }\perp
}^{2}+m^{2}-i\varepsilon }{k_{1^{\prime }}^{+}}\right) }\frac{1}{\left(
K^{-}-k_{1^{\prime }}^{-}-\frac{\left( K-k_{1^{\prime }}\right) _{\perp
}^{2}+m^{2}-i\varepsilon }{K^{+}-k_{1^{\prime }}^{+}}\right) }\times  \notag
\\
&&\frac{1}{\left( k_{2}^{-}-\frac{k_{2\perp }^{2}+m^{2}-i\varepsilon }{%
k_{2}^{+}}\right) }\frac{1}{\left( K^{-}-k_{2}^{-}-\frac{\left(
K-k_{2}\right) _{\perp }^{2}+m^{2}-i\varepsilon }{K^{+}-k_{2}^{+}}\right) }\
.  \label{ds2}
\end{eqnarray}

The double integration in $k^{-}$ is performed analytically in Eq.(\ref{ds2}%
). To simplify the notation, we define $q\equiv k_{1^{\prime }}$ and $%
k\equiv k_{1}$ . The integration is nonzero for $K^{+}>k^{+}>0$ and $%
K^{+}>q^{+}>0$. Two possibilities also appear for the forward propagation of 
$\sigma $. For $k^{+}>q^{+}$ it is created by particle $1$ and for $%
k^{+}<q^{+}$ it is annihilated by particle $1$: 
\begin{eqnarray}
\Delta {S}_{g_{S}^{2}}^{(2)}(K^{-}) &=&\left( ig_{S}\right) ^{2}\frac{\theta
(q^{+})\theta (K^{+}-q^{+})}{q^{+}\left( K^{+}-q^{+}\right) }\frac{i}{\left(
K^{-}-K_{0}^{^{\prime }(2)-}+i\varepsilon \right) }\times  \notag \\
&&\left( \frac{\theta (k^{+}-q^{+})}{\left( k^{+}-q^{+}\right) }\frac{i}{%
\left( K^{-}-K_{0}^{I(3)-}+i\varepsilon \right) }+\left[ k\leftrightarrow q%
\right] \right) \times  \notag \\
&&\frac{\theta (k^{+})\theta (K^{+}-k^{+})}{k^{+}\left( K^{+}-k^{+}\right) }%
\frac{i}{\left( K^{-}-K_{0}^{(2)-}+i\varepsilon \right) }\ ,  \label{corrg2}
\end{eqnarray}
where the light-front energies of the intermediate state propagation are
given by Eq.(\ref{k02}) for the initial and final two particle intermediate
states and by $K_{0}^{I(3)-}$ which is given by Eq.(\ref{k0n}). We have 
\begin{eqnarray}
K_{0}^{^{\prime }(2)-} &=&\frac{q_{\perp }^{2}+m^{2}}{q^{+}}+\frac{\left(
K-q\right) _{\bot }^{2}+m^{2}}{\left( K^{+}-q^{+}\right) },  \notag \\
K_{0}^{I(3)-} &=&\frac{q_{\perp }^{2}+m^{2}}{q^{+}}+\frac{\left( K-k\right)
_{\bot }^{2}+m^{2}}{\left( K^{+}-k^{+}\right) }+\frac{k_{\sigma \perp
}^{2}+\mu ^{2}}{k_{\sigma }^{+}},  \notag \\
K_{0}^{(2)-} &=&\frac{k_{\perp }^{2}+m^{2}}{k^{+}}+\frac{\left( K-k\right)
_{\bot }^{2}+m^{2}}{\left( K^{+}-k^{+}\right) },
\end{eqnarray}
where $k_{\sigma }^{+}=k^{+}-q^{+}$ and $\vec{k}_{\sigma \perp }=\vec{k}%
_{\perp }-\vec{q}_{\perp }$ .

The matrix elements of the interaction Hamiltonian that creates or destroys
a quantum of the intermediate boson are given by 
\begin{eqnarray}
<qk_{\sigma }|V|k> &=&2\delta (q+k_{\sigma }-k)\frac{g}{\sqrt{q^{+}k_{\sigma
}^{+}k^{+}}}\theta (k_{\sigma }^{+})  \notag \\
<q|V|k_{\sigma }k> &=&2\delta (k+k_{\sigma }-q)\frac{g}{\sqrt{q^{+}k_{\sigma
}^{+}k^{+}}}\theta (k_{\sigma }^{+})\ .  \label{v}
\end{eqnarray}

The perturbative correction of the propagator can be written as a
perturbative correction to the two-particle Green's function. Using Eqs. (%
\ref{g02}), (\ref{g0n}) and (\ref{v}), we have: 
\begin{equation}
\Delta
G_{g_{S}^{2}}^{(2)}(K^{-})=G_{0}^{(2)}(K^{-})VG_{0}^{(3)}(K^{-})VG_{0}^{(2)}(K^{-})\ .
\label{dgreen2}
\end{equation}
The correction $\Delta G_{g_{S}^{2}}^{(2)}(K^{-})$ also contains the self
energies diagrams for the bosons $\Phi _{1}$ and $\Phi _{2}$. Imposing the
restriction that only ladder diagrams are allowed in Eq.(\ref{dgreen2}), we
recover Eq.(\ref{corrg2}).

The nonpertubative Green's function up to order $g_{S}^{2}$ also contains
the covariant propagation in the light-front time up to one boson exchange,
which is written as: 
\begin{equation}
G_{g_{S}^{2}}^{(2)}(K^{-})=G_{0}^{(2)}(K^{-})+G_{0}^{(2)}(K^{-})K_{I}^{(3)-}G_{g_{S}^{2}}^{(2)}(K^{-})\ ,
\label{green2}
\end{equation}
where the interaction is 
\begin{equation}
K_{I}^{(3)-}=VG_{0}^{(3)}(K^{-})V\ .  \label{k3i}
\end{equation}

The above Green's function up to order $g_{S}^{2}$ reduces to the covariant
propagator for the light-front time. Restricting the kernel to the ladder
aproximation, 
\begin{equation}
K_{I,l}^{(3)-}=\left[ VG_{0}^{(3)}(K^{-})V\right] _{\text{ladder}}\ ,
\label{k3ladder}
\end{equation}
the bound state solutions of Eq.(\ref{green2}) satisfy the Weinberg equation 
\cite{wein}, which has already been solved numerically \cite{ji86}. Without
the ladder restriction, the bound state solution of Eq.(\ref{green2}) has
been obtained numerically in Ref.\cite{ji94}.

\subsection{Green's Function $\mathcal{O}(g_S^4)$}

The perturbative correction to the Feynman two-body propagator of order $%
g_{S}^{4}$ in the ladder approximation (box diagram) is given by: 
\begin{eqnarray}
\Delta S_{g_{S}^{4}}^{(2)}(x^{+}) &=&(\frac{ig_{S}}{2})^{4}\int d\overline{x}%
_{1}^{^{\prime }+}d\overline{x}_{2}^{^{\prime }+}d\overline{x}_{1}^{+}d%
\overline{x}_{2}^{+}\widetilde{S}_{1^{\prime }}^{(1)}(x^{+}-\overline{x}%
_{1}^{^{\prime }+})\widetilde{S}_{2^{\prime }}^{(1)}(x^{+}-\overline{x}%
_{2}^{^{\prime }+})\widetilde{S}_{\sigma ^{\prime }}^{(1)}(\overline{x}%
_{1}^{^{\prime }+}-\overline{x}_{2}^{^{\prime }+})\times  \notag \\
&&\widetilde{S}_{\overline{1}}^{(1)}(\overline{x}_{1}^{^{\prime }+}-%
\overline{x}_{1}^{+})\widetilde{S}_{\overline{2}}^{(1)}(\overline{x}%
_{2}^{^{\prime }+}-\overline{x}_{2}^{+})\ \widetilde{S}_{\sigma }^{(1)}(%
\overline{x}_{1}^{+}-\overline{x}_{2}^{+})\widetilde{S}_{1}^{(1)}(\overline{x%
}_{1}^{+})\widetilde{S}_{2}^{(1)}(\overline{x}_{2}^{+})\ .  \label{st2g4}
\end{eqnarray}

Writing the Fourier transform of the perturbative correction to the
propagator for $K^+>0$, we have: 
\begin{eqnarray}
&&\Delta {S}^{(2)}_{g_S^4}(K^{-}) =\frac{(ig_S)^4}{(2\pi)^3} \int \frac{%
dk^-dp^-dq^-} {k^+p^+q^+\left( K^+-k^+\right)\left(
K^+-p^+\right)\left(K^+-q^+\right) \left( q^+-p^+\right)\left( k^+-p^+\right)%
}\times  \notag \\
&&\frac 1{\left( q^--\frac{q_\perp ^2+m^2-i\varepsilon }{q^+}\right)} \frac
1{\left( K^--q^--\frac{\left( K-q\right) _\perp^2+ m^2-i\varepsilon }{K^+-q^+%
}\right) } \frac 1{\left( q^--p^--\frac{\left( q-p\right)
_\perp^2+m_\sigma^2 -i\varepsilon }{ q^+-p^+}\right) }\times  \notag \\
&&\frac 1{\left( p^{-}-\frac{p_{\perp }^2+m^2-i\varepsilon }{p^{+}}\right) }
\frac 1{\left(K^--p^--\frac{\left( K-p\right)_\perp ^2+m^2-i\varepsilon } {%
K^+-p^+}\right) } \frac{1}{\left( k^--p^-- \frac{\left( k-p \right)_\perp
^2+m_\sigma ^2-i\varepsilon }{k^+-p^+ }\right)} \times  \notag \\
&&\frac 1{\left( k^{-}-\frac{k_{\perp }^2+m^2-i\varepsilon }{k^{+}}\right)}
\frac 1{\left( K^{-}-k^{-}-\frac{\left( K-k\right)_{\perp}^2
+m^2-i\varepsilon }{ K^+-k^+}\right) } \ ,  \label{s2g4}
\end{eqnarray}
where we have defined $q\equiv k_{1^{\prime}}$, $p\equiv k_{\overline 1}$
and $k\equiv k_1$ to simplify the notation.

The correction to the propagator is found by analytical integration in the
light-front energies in Eq.(\ref{s2g4}). To separate the intermediate four
particle propagation, that occurs for $k^{+},\ p^{+}$ and $q^{+}$ such that $%
0<k^{+}<p^{+}<q^{+}<K^{+}$, the following factorization is necessary 
\begin{eqnarray}
&&\frac{1}{K^{-}-p^{-}-\frac{\left( K-p\right) _{\perp
}^{2}+m^{2}-i\varepsilon }{K^{+}-p^{+}}}\times \frac{1}{p^{-}-k^{-}-\frac{%
\left( k-p\right) _{\perp }^{2}+m_{\sigma }^{2}-i\varepsilon }{p^{+}-k^{+}}}
\notag \\
&=&\frac{1}{K^{-}-k^{-}-\frac{\left( K-p\right) _{\perp
}^{2}+m^{2}-i\varepsilon }{K^{+}-p^{+}}-\frac{\left( k-p\right) _{\perp
}^{2}+m_{\sigma }^{2}-i\varepsilon }{p^{+}-k^{+}}}\times  \notag \\
&&\left[ \frac{1}{K^{-}-p^{-}-\frac{\left( K-p\right) _{\perp
}^{2}+m^{2}-i\varepsilon }{K^{+}-p^{+}}}+\frac{1}{p^{-}-k^{-}-\frac{\left(
k-p\right) _{\perp }^{2}+m_{\sigma }^{2}-i\varepsilon }{p^{+}-k^{+}}}\right]
\ .
\end{eqnarray}

After the Cauchy integration the result for the correction to the two body
propagator, with the condition that $0<k^{+}<p^{+}<q^{+}<K^{+}$ $\left(
\Delta _{a}{S}_{g_{S}^{4}}^{(2)}\right) $, is given by 
\begin{eqnarray}
\Delta _{a}{S}_{g_{S}^{4}}^{(2)}(K^{-}) &=&(ig_{S})^{4}\frac{\theta
(k^{+})\theta (K^{+}-k^{+})}{k^{+}(K^{+}-k^{+})}\frac{i}{K^{-}-\frac{%
k_{\perp }^{2}+m^{2}-i\varepsilon }{k^{+}}-\frac{\left( K-k\right) _{\perp
}^{2}+m^{2}-i\varepsilon }{K^{+}-k^{+}}}\times  \notag \\
&&\frac{\theta (p^{+})\theta (q^{+}-p^{+})\theta (p^{+}-k^{+})}{%
(q^{+}-p^{+})(p^{+}-k^{+})(K^{+}-p^{+})p^{+}}\left[ {F}^{\prime }(K^{-})+{F}%
^{^{\prime \prime }}(K^{-})\right] \times  \notag \\
&&\frac{\theta (q^{+})\theta (K^{+}-q^{+})}{q^{+}(K-q)^{+}}\frac{i}{K^{-}-%
\frac{q_{\perp }^{2}+m^{2}-i\varepsilon }{q^{+}}-\frac{\left( K-q\right)
_{\perp }^{2}+m^{2}-i\varepsilon }{K^{+}-q^{+}}}\ ,  \label{delta}
\end{eqnarray}
with 
\begin{eqnarray}
{F}^{\prime }(K^{-}) &=&\frac{i}{K^{-}-\frac{k_{\perp }^{2}+m^{2}}{k^{+}}-%
\frac{\left( K-p\right) _{\perp }^{2}+m^{2}}{K^{+}-p^{+}}-\frac{\left(
p-k\right) _{\perp }^{2}+m^{2}}{p^{+}-k^{+}}}\times  \notag \\
&&\frac{i}{K^{-}-\frac{p_{\perp }^{2}+m^{2}}{p^{+}}-\frac{\left( K-q\right)
_{\perp }^{2}+m^{2}}{K^{+}-q^{+}}-\frac{\left( q-p\right) _{\perp }^{2}+m^{2}%
}{q^{+}-p^{+}}}\times  \notag \\
&&\frac{i}{K^{-}-\frac{p_{\perp }^{2}+m^{2}}{p^{+}}-\frac{\left( K-p\right)
_{\perp }^{2}+m^{2}}{K^{+}-p^{+}}}\ ;  \label{fl}
\end{eqnarray}
\begin{eqnarray}
{F}^{\prime \prime }(K^{-}) &=&\frac{i}{K^{-}-\frac{k_{\perp }^{2}+m^{2}}{%
k^{+}}-\frac{\left( K-p\right) _{\perp }^{2}+m^{2}}{K^{+}-p^{+}}-\frac{%
\left( p-k\right) _{\perp }^{2}+m^{2}}{p^{+}-k^{+}}}\times  \notag \\
&&\frac{i}{K^{-}-\frac{p_{\perp }^{2}+m^{2}}{p^{+}}-\frac{\left( K-q\right)
_{\perp }^{2}+m^{2}}{K^{+}-q^{+}}-\frac{\left( q-p\right) _{\perp
}^{2}+m_{\sigma }^{2}}{q^{+}-p^{+}}}\times  \notag \\
&&\frac{i}{K^{-}-\frac{k_{\perp }^{2}+m^{2}}{k^{+}}-\frac{\left( K-q\right)
_{\perp }^{2}+m^{2}}{K^{+}-q^{+}}-\frac{\left( q-p\right) _{\perp
}^{2}+m_{\sigma }^{2}}{q^{+}-p^{+}}-\frac{\left( p-k\right) _{\perp
}^{2}+m_{\sigma }^{2}}{p^{+}-k^{+}}}.  \label{fll}
\end{eqnarray}

The part of the propagator given by Eq.(\ref{delta}) contains the virtual
light-front propagation of intermediate states with up to four particles.
The function $F^{\prime }$ contains only intermediate states up to three
particles, while $F^{\prime \prime }$ has one intermediate state propagation
with four particles which can be recognized as the last term of Eq.(\ref{fll}%
). The other possibility that includes up to four particles in the
intermediate state propagation is given by $0<q^{+}<p^{+}<k^{+}<K^{+}$. To
obtain this contribution, we make the transformation $q\leftrightarrow k$ in
Eq.(\ref{delta}).

The correction to the propagator $\left( \Delta _{b}{S}_{g_{S}^{4}}^{(2)}%
\right) $ for $0<p^{+}<k^{+}<q^{+}<K^{+}$ contains only up to three-particle
intermediate states only and is given by 
\begin{eqnarray}
&&\Delta _{b}{S}_{g_{S}^{4}}^{(2)}(K^{-})=(ig_{S})^{4}\frac{\theta
(k^{+})\theta (K^{+}-k^{+})}{k^{+}(K-k)^{+}}\frac{i}{K^{-}-\frac{k_{\perp
}^{2}+m^{2}-i\varepsilon }{k^{+}}-\frac{\left( K-k\right) _{\perp
}^{2}+m^{2}-i\varepsilon }{K^{+}-k^{+}}}\times  \notag \\
&&\frac{\theta (q^{+}-k^{+})\theta (k^{+}-p^{+})}{%
(q^{+}-p^{+})(k^{+}-p^{+})(K^{+}-p^{+})p^{+}}\frac{i}{K^{-}-\frac{p_{\perp
}^{2}+m^{2}}{p^{+}}-\frac{\left( K-q\right) _{\perp }^{2}+m^{2}}{K^{+}-q^{+}}%
-\frac{\left( q-p\right) _{\perp }^{2}+m_{\sigma }^{2}}{q^{+}-p^{+}}}\times 
\notag \\
&&\frac{i}{K^{-}-\frac{p_{\perp }^{2}+m^{2}}{p^{+}}-\frac{\left( K-p\right)
_{\perp }^{2}+m^{2}}{K^{+}-p^{+}}}\frac{i}{K^{-}-\frac{p_{\perp }^{2}+m^{2}}{%
p^{+}}-\frac{\left( K-k\right) _{\perp }^{2}+m^{2}}{K^{+}-k^{+}}-\frac{%
\left( k-p\right) _{\perp }^{2}+m_{\sigma }^{2}}{k^{+}-p^{+}}}\times  \notag
\\
&&\frac{\theta (q^{+})\theta (k^{+}-q^{+})}{q^{+}(K^{+}-q^{+})}\frac{i}{%
K^{-}-\frac{q_{\perp }^{2}+m^{2}-i\varepsilon }{q^{+}}-\frac{\left(
K-q\right) _{\perp }^{2}+m^{2}-i\varepsilon }{K^{+}-q^{+}}}\ .  \label{deltb}
\end{eqnarray}

For the momenta satisfying $0<q^{+}<k^{+}<p^{+}<K^{+}$, the correction to
the propagator can be obtained from Eq.(\ref{deltb}) by performing the
transformation $q\leftrightarrow K-q$ and $k\leftrightarrow K-k$. From Eqs.(%
\ref{delta}) and (\ref{deltb}), the following result is obtained 
\begin{eqnarray}
\Delta {S}_{g_{S}^{4}}^{(2)}(K^{-}) &=&\Delta _{a}{S}%
_{g_{S}^{4}}^{(2)}(K^{-})+\Delta _{a}{S}_{g_{S}^{4}}^{(2)}(K^{-})\left[
q\leftrightarrow k\right] +  \notag \\
&&\Delta _{b}{S}_{g_{S}^{4}}^{(2)}(K^{-})+\Delta _{b}{S}%
_{g_{S}^{4}}^{(2)}(K^{-})\left[ q\leftrightarrow K-q,\ k\leftrightarrow K-k%
\right] \ .  \label{deltab}
\end{eqnarray}

Finally, the correction in order $g_{S}^{4}$ to the Green's function can be
found identifying the free Green's functions for two, three and four body
propagation in Eqs.(\ref{delta}) and (\ref{deltb}). In its general form, 
\begin{eqnarray}
\Delta G_{g_{S}^{4}}^{(2)}(K^{-})
&=&G_{0}^{(2)}(K^{-})VG_{0}^{(3)}(K^{-})VG_{0}^{(2)}(K^{-})VG_{0}^{(3)}(K^{-})VG_{0}^{(2)}(K^{-})+
\notag \\
&&G_{0}^{(2)}(K^{-})VG_{0}^{(3)}(K^{-})VG_{0}^{(4)}(K^{-})VG_{0}^{(3)}(K^{-})VG_{0}^{(2)}(K^{-})\ ,
\label{dgreen4}
\end{eqnarray}
it contains the self energies corrections for the bosons $\Phi _{1}$ and $%
\Phi _{2}$, vertex corrections and the crossed box diagram. Imposing the
restriction that only ladder diagrams are allowed in Eq.(\ref{dgreen4}), we
recover Eq.(\ref{s2g4}).

The nonpertubative Green's function up to order $g_{S}^{4}$ contains the
covariant propagation of the system in the ladder approximation and is
written as: 
\begin{equation}
G_{g_{S}^{4}}^{(2)}(K^{-})=G_{0}^{(2)}(K^{-})+G_{0}^{(2)}(K^{-})\left(
K_{I}^{(3)-}+K_{I}^{(4)-}\right) G_{g_{S}^{4}}^{(2)}(K^{-})\ ,
\label{greeng4}
\end{equation}
where 
\begin{equation}
K_{I}^{(4)-}=VG_{0}^{(3)}(K^{-})VG_{0}^{(4)}(K^{-})VG_{0}^{(3)}(K^{-})V\ .
\end{equation}
The Green's function, (\ref{greeng4}), up to order $g_{S}^{2}$ yields the
covariant propagator for light-front times. Separating the terms in the
kernel corresponding to the covariant ladder, 
\begin{equation}
K_{I,l}^{(4)-}=\left[
VG_{0}^{(3)}(K^{-})VG_{0}^{(4)}(K^{-})VG_{0}^{(3)}(K^{-})V\right] _{\text{%
ladder}}\ ,  \label{k4ladder}
\end{equation}
together with the contribution of the one boson exchange diagram, Eq.(\ref%
{k3ladder}), we obtain the nonperturbative Green's function equation which,
up to order $g_{S}^{4}$, corresponds to the covariant ladder propagator for
light-front times, including one boson exchange and the box diagram. It is
given by 
\begin{equation}
G_{g_{S}^{4},l}^{(2)}(K^{-})=G_{0}^{(2)}(K^{-})+G_{0}^{(2)}(K^{-})\left(
K_{I,l}^{(3)-}+K_{I,l}^{(4)-}\right) G_{g_{S}^{4},l}^{(2)}(K^{-})\ ,
\label{g41}
\end{equation}
in which the bound state solutions of Eq.(\ref{g41}) have up to four
particles in lowest order in the intermediate state. We have solved it
numerically to study the contribution of this Fock component in the
nonperturbative regime in which the bound state appears \cite{00sales}.

\subsection{Green's Function $\mathcal{O}(g_S^{2n})$}

The perturbative correction to the light-front propagator of order $%
g_{S}^{2n}$ for the covariant ladder can be obtained by generalizing our
previous results as 
\begin{eqnarray}
\Delta {S}_{g_{S}^{2n}}^{(2)}(K^{-}) &=&i^{3n+2}(ig_{S})^{2n}\int \frac{%
dk_{1}^{-}}{k_{1}^{+}(K^{+}-k_{1}^{+})}\frac{1}{\left( k_{1}^{-}-\frac{%
k_{1\perp }^{2}+m^{2}-i\varepsilon }{k_{1}^{+}}\right) }\frac{1}{\left(
K^{-}-k_{1}^{-}-\frac{\left( K-k_{1}\right) _{\perp }^{2}+m^{2}-i\varepsilon 
}{K^{+}-k_{1}^{+}}\right) }\times  \notag \\
&&\prod_{i=2}^{n}\frac{dk_{i}^{-}}{%
k_{i}^{+}(K^{+}-k_{i}^{+})(k_{i}^{+}-k_{i-1}^{+})}\frac{1}{\left(
k_{i}^{-}-k_{i-1}^{-}-\frac{\left( k_{i}-k_{i-1}\right) _{\perp
}^{2}+m_{\sigma }^{2}-i\varepsilon }{k_{i}^{+}-k_{i-1}^{+}}\right) }\times 
\notag \\
&&\frac{1}{\left( k_{i}^{-}-\frac{k_{i\perp }^{2}+m^{2}-i\varepsilon }{%
k_{i}^{+}}\right) }\frac{1}{\left( K^{-}-k_{i}^{-}-\frac{\left(
K-k_{i}\right) _{\perp }^{2}+m^{2}-i\varepsilon }{K^{+}-k_{i}^{+}}\right) }\
.  \label{dsg2n}
\end{eqnarray}
In Eq.(\ref{dsg2n}), light-front intermediate states of up to $n+2$
particles appear. %The terms that have less than $n+2$ 
%virtual particles in the intermediate state propagation
%can be decomposed into iterations of the  Green's function 
%in lowest order of $g_S$.

Formally we can write the light-front propagator up to order $g_S^{2n}$ as a
sum given by: 
\begin{eqnarray}
{S}^{(2)}_{g_S^{2n}}(K^{-})={S}^{(2)}(K^{-})+ \sum_{m=1}^n\Delta {S}%
^{(2)}_{g_S^{2m}}(K^{-}) \ .  \label{s2gn}
\end{eqnarray}

\section{Hierarchy equations}

The light-front Green's function for the two-body system obtained from the
solution of the covariant BS equation that contains all two-body irreducible
diagrams, with the exception of those including 
%self energy corrections of the intermediate boson, and amplitudes 
%corresponding to the creation or absorption of bosons $\sigma$ by 
%internal 
closed loops of bosons $\Phi _{1}$ and $\Phi _{2}$ and part of the
cross-ladder diagrams, is given by: 
\begin{eqnarray}
G^{(2)}(K^{-})
&=&G_{0}^{(2)}(K^{-})+G_{0}^{(2)}(K^{-})VG^{(3)}(K^{-})VG^{(2)}(K^{-})\ , 
\notag \\
G^{(3)}(K^{-})
&=&G_{0}^{(3)}(K^{-})+G_{0}^{(3)}(K^{-})VG^{(4)}(K^{-})VG^{(3)}(K^{-})\ , 
\notag \\
G^{(4)}(K^{-})
&=&G_{0}^{(4)}(K^{-})+G_{0}^{(4)}(K^{-})VG^{(5)}(K^{-})VG^{(4)}(K^{-})\ , 
\notag \\
&&\ .\ .\ .  \notag \\
G^{(N)}(K^{-})
&=&G_{0}^{(N)}(K^{-})+G_{0}^{(N)}(K^{-})VG^{(N+1)}(K^{-})VG^{(N)}(K^{-})\ , 
\notag \\
&&\ .\ .\ .  \label{coupled}
\end{eqnarray}
The hierarchy of equations (\ref{coupled}) corresponds to a truncation in
the light-front Fock space in which only two bosons states with the two
particles $\Phi _{1}$ and $\Phi _{2}$ are allowed in the intermediate state,
with no any restriction on the number of bosons $\sigma $, which thus
excludes the complete representation of the crossed ladder diagrams. To
obtain the two-body propagator for light-front times in the covariant ladder
approximation, the kernel of the hierarchy equations must be restricted.

A systematic expansion by the consistent truncation of the light-front Fock
space up to $N$ particles in the intermediate states (boson 1, boson 2 and $%
N-2$ $\sigma $'s) in the set of Eqs.(\ref{coupled}), amounts to substitution 
\begin{equation}
G^{(N)}(K^{-})\rightarrow G_{0}^{(N)}(K^{-})\ ,  \label{truncn}
\end{equation}
and subsequent solution of the coupled-hierarchy equations.

By restricting to up to four-particles in the intermediate state
propagation, we obtain the following nonperturbative equation for the
Green's function: 
\begin{eqnarray}
G^{(2)}(K^{-})
&=&G_{0}^{(2)}(K^{-})+G_{0}^{(2)}(K^{-})VG^{(3)}(K^{-})VG^{(2)}(K^{-})\ ,
\label{g24} \\
G^{(3)}(K^{-})
&=&G_{0}^{(3)}(K^{-})+G_{0}^{(3)}(K^{-})VG_{0}^{(4)}(K^{-})VG^{(3)}(K^{-})\ .
\label{g34}
\end{eqnarray}
The kernel of Eq.(\ref{g24}) still contains an infinite sum of light-front
diagrams, that are obtained solving by Eq.(\ref{g34}). To obtain the ladder
aproximation up to order $g_{S}^{4}$, Eq.(\ref{g41}), only the free and
first order terms are kept in Eq.(\ref{g34}), with the restriction of only
one and two boson covariant exchanges.

The bound-state solution of Eq.(\ref{coupled}) is covariant in the sense
that only the initial and final states are defined for specific light-front
times, and thus the position of the bound state pole is frame independent.
The residue of $G^{(2)}(K^{-})$ at the bound state energy $K^{-}=K_{B}^{-}$
is related to the two-body wave-function at given light-front time.

The two-body T-Matrix is written in terms of the light-front two-body
Green's function, Eq.(\ref{coupled}), as 
\begin{equation}
T^{(2)}(K^{-})=VG^{(3)}(K^{-})V+VG^{(3)}(K^{-})VG^{(2)}(K^{-})VG^{(3)}(K^{-})V\ .
\label{tmatrix}
\end{equation}
The T-matrix also satisfies the inhomogeneous integral equation: 
\begin{equation}
T^{(2)}(K^{-})=VG^{(3)}(K^{-})V+VG^{(3)}(K^{-})VG_{0}^{(2)}(K^{-})T^{(2)}(K^{-})\ .
\label{eqtmatrix}
\end{equation}
Note that the T-matrix equation above is exact in providing the
non-perturbative two-body propagator given by the Green's function $%
G^{(2)}(K^{-})$. The interaction that appears in Eq.(\ref{eqtmatrix}) has an
infinite number of terms, as one sees in the second equation of the
hierarchy Eqs.(\ref{coupled}). By truncating $G^{(3)}(K^{-})$ it is possible
to define several approximate equations. In particular the choice of $%
G_{0}^{(3)}(K^{-})$ to approximate $G^{(3)}(K^{-})$, gives the bound state
equation found by Weinberg \cite{wein}.

The covariant on-shell scattering amplitude, $t^{(2)}(K^{-})$, is obtained
from the T-matrix, by taking into account the phase space factors: 
\begin{equation}
<q,K-q|t^{(2)}(K^{-})|k,K-k>=\frac{<q,K-q|T^{(2)}(K^{-})|k,K-k>}{\left[
q^{+}(K^{+}-q^{+})k^{+}(K^{+}-k^{+})\right] ^{\frac{1}{2}}}\ .
\end{equation}

The full-off-shell covariant two-body amplitude, is formally obtained from
the light-front Green's function, Eq.(\ref{coupled}), by substituting the
values of the on-shell initial and final ``$-$'' momentum by their off-shell
values 
\begin{eqnarray}
&<&q,K-q|t_{(off)}^{(2)}(K^{-})|k,K-k>=  \notag \\
&&\underset{(off)}{}\left[ \frac{%
VG^{(3)}(K^{-})V+VG^{(3)}(K^{-})VG^{(2)}(K^{-})VG^{(3)}(K^{-})V}{\left[
q^{+}(K^{+}-q^{+})k^{+}(K^{+}-k^{+})\right] ^{\frac{1}{2}}}\right] _{(off)};
\label{tmatrixoff}
\end{eqnarray}
such that the initial off-energy-shell momenta are $k^{-}$ and $K^{-}-k^{-}$
and the final off-energy-shell ones are $q^{-}$ and $K^{-}-q^{-}$.

\section{Light-Front Bound state Equation}

The homogeneous equation for the light-front two-body bound state
wave-function is obtained the solution of 
\begin{equation}
|\Psi _{B}>=G_{0}^{(2)}(K_{B}^{-})VG^{(3)}(K_{B}^{-})V|\Psi _{B}>\ ,
\end{equation}
with the kernel defined by the hierarchy Eqs.(\ref{coupled}). It can also be
written as an eigenvalue equation for the squared mass operator: 
\begin{equation}
\left[ \left( M_{0}^{(2)}\right) ^{2}+K^{+}VG^{(3)}(K_{B}^{-})V\right] |\Psi
_{B}>=(M_{2})^{2}|\Psi _{B}>,
\end{equation}
where $(M_{2})^{2}=K^{+}K_{B}^{-}-K_{\perp }^{2}$ and $%
M_{0}^{(2)}=K^{+}K_{0}^{(2)-}-K_{\perp }^{2}$. %Eq.(\ref{k02}).

The vertex function for the bound state wave-function is defined as 
\begin{equation}
\Gamma _{LF}(\vec{q}_{\perp },q^{+})=<q,\ K-q|\left(
G_{0}^{(2)}(K_{B}^{-})\right) ^{-1}|\Psi _{B}>.  \label{vertexlf}
\end{equation}

The light-front wave-function at $x^{+}=0$ is obtained from the residue of
the covariant two-body propagator at the bound state pole: 
\begin{eqnarray}
G_{0}^{(2)}(K_{B}^{-})\frac{\Gamma _{LF}(\vec{q}_{\perp },q^{+})}{\sqrt{%
q^{+}(K^{+}-q^{+})}} &=&\frac{1}{\sqrt{q^{+}(K^{+}-q^{+})}}<q,\ P-q|\Psi
_{B}>  \notag \\
&=&\int dq^{-}\frac{F(q^{\mu })}{(q^{2}-m^{2}+i\varepsilon
)((k-q)^{2}-m^{2}+i\varepsilon )}\ .  \label{lfwf}
\end{eqnarray}

Approaching the bound-state pole of the off-shell T-matrix, Eq. (\ref%
{tmatrixoff}), we relate the residue at this pole to the full
four-dimensional vertex $F(q^{\mu })$, which is the solution of the
field-theoretical BS equation. In terms of the light-front wave-function,
the four-dimensional vertex is given by 
\begin{equation*}
F(q^{\mu })=\frac{1}{\sqrt{q^{+}(K^{+}-q^{+})}}<q,\ K-q|_{(off)}\left[
VG^{(3)}(K_{B}^{-})V\right] |\Psi _{B}>\ .
\end{equation*}
The operator $_{(off)}\left[ VG^{(3)}(K_{B}^{-})V\right] $ has off-shell
energy values for $q_{1}^{-}=q^{-}$ and $q_{2}^{-}=K_{B}^{-}-q^{-}$ in the
positions where the corresponding on-shell values appear in the many-body
Green's function.

We redefine the light-front vertex function by includins the phase-space
factor, in order to simplify the formula for the bound state equation: $%
F_{LF}=\sqrt{q^{+}(K^{+}-q^{+})}\Gamma _{LF}$.

In the actual numerical calculations to quantify the effects of the higher
Fock-components, we have used the bound state solution of Eq.(\ref{g41}) 
\cite{00sales}. The Green's function obtained from this equation, up to
order $g_{S}^{4}$, reproduces the covariant two-body propagator between two
light-front hypersurfaces. In this approximation, the vertex function
satisfies the following integral equation, 
\begin{equation}
F_{LF}(\vec{q}_{\bot },y)=\frac{1}{(2\pi )^{3}}\int \frac{d^{2}k_{\bot }dx}{%
2x(1-x)}\frac{\overline{K}_{I,l}^{(3)-}(\vec{q}_{\bot },y;\vec{k}_{\bot },x)+%
\overline{K}_{I,l}^{(4)-}(\vec{q}_{\bot },y;\vec{k}_{\bot },x)}{%
M_{2}^{2}-M_{0}^{2}}F_{FL}(\vec{k}_{\bot },x)\ ,  \label{bsnp}
\end{equation}
where the momentum fractions are $y=q^{+}/K^{+}$ and $x=k^{+}/K^{+}$, with $%
0<y<1$.

The part of the kernel which contains only the propagation of virtual three
particle states foward in the light-front time is obtained from Eq.(\ref%
{k3ladder}) as, 
\begin{eqnarray}
&&\overline{K}_{I,l}^{\left( 3\right) -}(\vec{q}_{\bot },y;\vec{k}_{\bot
},x)=\sqrt{\frac{q^{+}(K^{+}-q^{+})}{k^{+}(K^{+}-k^{+})}}\left[
K_{I,l}^{\left( 3\right) -}(\vec{q}_{\bot },q^{+};\vec{k}_{\bot },k^{+})%
\right] =  \notag \\
&&g_{S}^{2}\frac{\theta (x-y)}{\left( x-y\right) \left( M_{2}^{2}-\frac{%
q_{\bot }^{2}+m^{2}}{y}-\frac{k_{\bot }^{2}+m^{2}}{1-x}-\frac{(\vec{q}_{\bot
}-\vec{k}_{\bot })^{2}+\mu ^{2}}{x-y}\right) }+\left[ x\leftrightarrow y,%
\vec{k}_{\bot }\leftrightarrow \vec{q}_{\bot }\right] .  \label{k3}
\end{eqnarray}
where the momentum fractions are $y=q^{+}/K^{+}$ and $x=k^{+}/K^{+}$. The
denominator in Eq.(\ref{k3}), in the non-relativistic limit gives origin to
the Yukawa potential of range $\mu ^{-1}$.

The contribution to the kernel from the virtual four-body propagation is
obtained from Eq.(\ref{k4ladder}) as, 
\begin{eqnarray}
&&\overline{K}_{I,l}^{\left( 4\right) -}(\vec{q}_{\bot },y;\vec{k}_{\bot
},x)=\sqrt{\frac{q^{+}(K^{+}-q^{+})}{k^{+}(K^{+}-k^{+})}}\left[
K_{I,l}^{\left( 4\right) -}(\vec{q}_{\bot },q^{+};\vec{k}_{\bot },k^{+})%
\right] =  \notag \\
&&\frac{g_{S}^{4}}{(2\pi )^{3}}\int \frac{d^{2}p_{\bot }dz}{2z(1-z)\left(
z-x\right) (y-z)}\frac{\theta (z-y)\theta (x-z)}{\left( M_{2}^{2}-\frac{%
q_{\bot }^{2}+m^{2}}{y}-\frac{p_{\bot }^{2}+m^{2}}{1-z}-\frac{(\vec{q}_{\bot
}-\vec{p}_{\bot })^{2}+\mu ^{2}}{z-y}\right) }  \notag \\
&\times &\frac{1}{\left( M_{2}^{2}-\frac{q_{\bot }^{2}+m^{2}}{y}-\frac{%
k_{\bot }^{2}+m^{2}}{1-x}-\frac{(\vec{q}_{\bot }-\vec{p}_{\bot })^{2}+\mu
^{2}}{z-y}-\frac{(\vec{p}_{\bot }-\vec{k}_{\bot })^{2}+\mu ^{2}}{x-z}\right) 
}  \notag \\
&\times &\frac{1}{\left( M_{2}^{2}-\frac{p_{\bot }^{2}+m^{2}}{z}-\frac{%
k_{\bot }^{2}+m^{2}}{1-x}-\frac{(\vec{p}_{\bot }-\vec{k}_{\bot })^{2}+\mu
^{2}}{x-z}\right) }+\left[ x\leftrightarrow y,\vec{k}_{\bot }\leftrightarrow 
\vec{q}_{\bot }\right] \ .  \label{k4}
\end{eqnarray}

\section{Conclusion}

We have developed a general framework for constructing the light-front
two-body Green's function and have discussed the formal representation of
the covariant BS equation in the light-front. We have show how to obtain the
covariant off-shell T-matrix and vertex of the bound state from the
light-front quantities. Although our discussion has been performed in the
contex of a bosonic Lagrangean, it can be extended to general cases \cite%
{01sales}. We have found a hierarchy of coupled equations which gives the
exact two-body propagator in several cases, including the ladder
approximation. Truncation of the hierarchy can be performed consistently by
approximating the $N$-body Green's function with the free operator, which
implies that the $N$-body light-front intermediate state is accounted for in
lowest order of the two-body propagation.

\section*{Acknowledgments}

J.H.O. Sales is supported by FAPESP. B.V.Carlson and T.Frederico acknowledge
partial support from FAPESP and the CNPq. B.M. Pimentel thanks to CNPq and
FAPESP (grant 02/00222-9) for partial support. All acknowledge partial
support from CAPES/DAAD project 015/95. T.F. thanks the hospitality of the
Institute for Theoretical Physics, University of Hannover, where part of
this work has been performed.

\end{document}